\documentclass[
 reprint,
superscriptaddress,
longbibliography,
 amsmath,amssymb,
 aps,
prx,
]{revtex4-2}

\usepackage{graphicx}
\usepackage{dcolumn}
\usepackage{bm}

\usepackage{mhchem}
\usepackage{hyperref}
\hypersetup{
    colorlinks = true,
    citecolor = {blue},
    linkcolor = {blue},
    urlcolor = {black}
}

\begin{document}
\title{High thermoelectric power factor through topological flat bands}

\author{Fabian Garmroudi}
\email{fabian.garmroudi@tuwien.ac.at}
\affiliation{Institute of Solid State Physics, TU Wien, 1040 Vienna, Austria} 
\author{Illia Serhiienko}
\affiliation{International~Center~for~Materials~Nanoarchitectonics~(WPI-MANA),~National~Institute~for~Materials~Science,~Tsukuba~305-0044,~Japan}
\affiliation{University of Tsukuba, Tsukuba 305-8577, Japan}
\author{Simone Di Cataldo}
\affiliation{Institute of Solid State Physics, TU Wien, 1040 Vienna, Austria}
\affiliation{Dipartimento di Fisica, Sapienza University of Rome, Piazzale Aldo Moro 5, 00185 Roma, Italy}
\author{Michael Parzer}
\affiliation{Institute of Solid State Physics, TU Wien, 1040 Vienna, Austria}
\author{Alexander Riss}
\affiliation{Institute of Solid State Physics, TU Wien, 1040 Vienna, Austria}
\author{Matthias Grasser}
\affiliation{Institute of Solid State Physics, TU Wien, 1040 Vienna, Austria}
\author{Simon Stockinger}
\affiliation{Institute of Solid State Physics, TU Wien, 1040 Vienna, Austria}
\author{Sergii Khmelevskyi}
\affiliation{Vienna Scientific Cluster Research Center, TU Wien, 1040 Vienna, Austria}
\author{Kacper Pryga}
\affiliation{AGH University of Krakow, Faculty of Physics and Applied Computer Science, Aleja
Mickiewicza 30, 30-059 Krakow, Poland}
\author{Bartlomiej Wiendlocha}
\affiliation{AGH University of Krakow, Faculty of Physics and Applied Computer Science, Aleja
Mickiewicza 30, 30-059 Krakow, Poland}
\author{Karsten Held}
\affiliation{Institute of Solid State Physics, TU Wien, 1040 Vienna, Austria}
\author{Takao Mori}
\affiliation{International~Center~for~Materials~Nanoarchitectonics~(WPI-MANA),~National~Institute~for~Materials~Science,~Tsukuba~305-0044,~Japan}
\affiliation{University of Tsukuba, Tsukuba 305-8577, Japan}
\author{Ernst Bauer}
\affiliation{Institute of Solid State Physics, TU Wien, 1040 Vienna, Austria}
\author{Andrej Pustogow}
\affiliation{Institute of Solid State Physics, TU Wien, 1040 Vienna, Austria}

\begin{abstract}
Thermoelectric (TE) materials are useful for applications such as waste heat harvesting or efficient and targeted cooling. While various strategies towards superior thermoelectrics through a reduction of the lattice thermal conductivity have been developed, a path to enhance the power factor is pressing. Here, we report large power factors up to 5\,mW\,m$^{-1}$\,K$^{-2}$ at room temperature in the kagome metal \ce{Ni3In_{1-x}Sn_x}. This system is predicted to feature almost dispersionless flat bands in conjunction with highly dispersive Dirac-like bands in its electronic structure around the Fermi energy \ce{\textit{E}_{\text{F}}} [L. Ye \textit{et al.}, \href{https://doi.org/10.1038/s41567-023-02360-5}{Nature Physics (2024)}]. Within this study, we experimentally and theoretically showcase that tuning this flat band precisely below \ce{\textit{E}_{\text{F}}} by chemical doping $x$ boosts the Seebeck coefficient and power factor, as highly mobile charge carriers scatter into the flat-band states. 
Our work demonstrates the prospect of engineering extremely flat and highly dispersive bands towards the Fermi energy in kagome metals and introduces topological flat bands as a novel tuning knob for thermoelectrics. 
\end{abstract}

\maketitle
\section{Introduction}
Flat-band materials are of immense current interest as they promise a rich tapestry of emergent correlation phenomena and novel physics to be discovered \cite{regnault2022catalogue,checkelsky2024flat}. Certain frustrated geometries such as the dice, Lieb or kagome lattices are theoretically predicted to support \textit{completely flat} electronic bands, \textit{i.\,e.}, a quasi-infinitely degenerate set of quantum states arising from internal symmetries or local topology \cite{sutherland1986localization,bergman2008band,leykam2018artificial}. For instance, within a simple single-orbital tight-binding model for the kagome lattice, destructive quantum interference among electronic hopping pathways leads to perfectly flat bands alongside highly dispersive Dirac bands as evidenced to some degree, for instance, in the kagome metals \ce{Fe3Sn2} \cite{ye2018massive}, FeSn \cite{kang2020dirac} and CoSn \cite{kang2020topological}. Such flat-band states are believed to hold the key towards designing novel quantum phases of matter as the electron--electron interaction becomes decisive compared to the quenched kinetic energy \cite{tang2011high,liu2014exotic,rosa2024quantum}. Here, we recognize the prospect of such topological band structure features -- given that they can be properly tuned around the Fermi level -- for thermoelectric materials. Thermoelectrics have attained significant attention for their ability to directly interconvert thermal and electrical energy via the Seebeck and Peltier effects \cite{pecunia2023roadmap}. An efficient TE material ought to possess a large power factor $PF=S^2\,\sigma$ and a small thermal conductivity $\kappa$, summarized in the dimensionless figure of merit $zT=S^2\sigma\,\kappa^{-1}\,T$. Here, $S$ denotes the Seebeck coefficient (thermovoltage per temperature gradient), $\sigma$ the electrical, and $\kappa$ the thermal conductivity consisting of lattice $\kappa_l$ and electron contributions $\kappa_e$; the latter can be expressed by the Wiedemann-Franz (WF) law $\kappa_e=L\sigma T$ ($L$ being the Lorenz number). Ever since Abram Ioffe proposed semiconductors as efficient thermoelectrics \cite{ioffe1959semiconductor}, there has been a surge of investigations on numerous complex systems with the primary focus on reducing $\kappa_l$ via heavy element substitution \cite{mikami2012thermoelectric,chen2013effect}, nanostructuring \cite{snyder2008complex,
biswas2012high} or to look for systems with intrinsically small $\kappa_l$ \cite{nolas1999skutterudites,zhao2014ultralow,mark2023ultralow}. As a downside, crystallographic and chemical features yielding ultralow lattice thermal conductivity, such as anharmonicity, weak bonding, structural defects, and complex nanostructures, often go hand in hand with poor structural, mechanical, and thermal stability, also compromising carrier mobility.
 
The stringent requirement to reduce $\kappa_l$, however, does not apply to good metallic conductors. Indeed, metals often exhibit $\kappa_e \gg \kappa_l$, which means that $zT$ simplifies to $zT=S^2/L$ owing to the WF law \cite{garmroudi2023high}. Since $L$ only varies moderately with temperature and composition in metals and because $L$ can even be estimated from measurements of $S$ \cite{kim2015characterization}, this leaves only a single tuning parameter -- the Seebeck coefficient $S$ -- that has to be taken into account. Moreover, metallic systems frequently comprise other desirable properties with respect to applications such as ductility and mechanical strength.

\begin{figure*}[t!]
\newcommand{\setwidth}{0.45}
			\centering
			\hspace*{0.25cm}
		\includegraphics[width=0.85\textwidth]{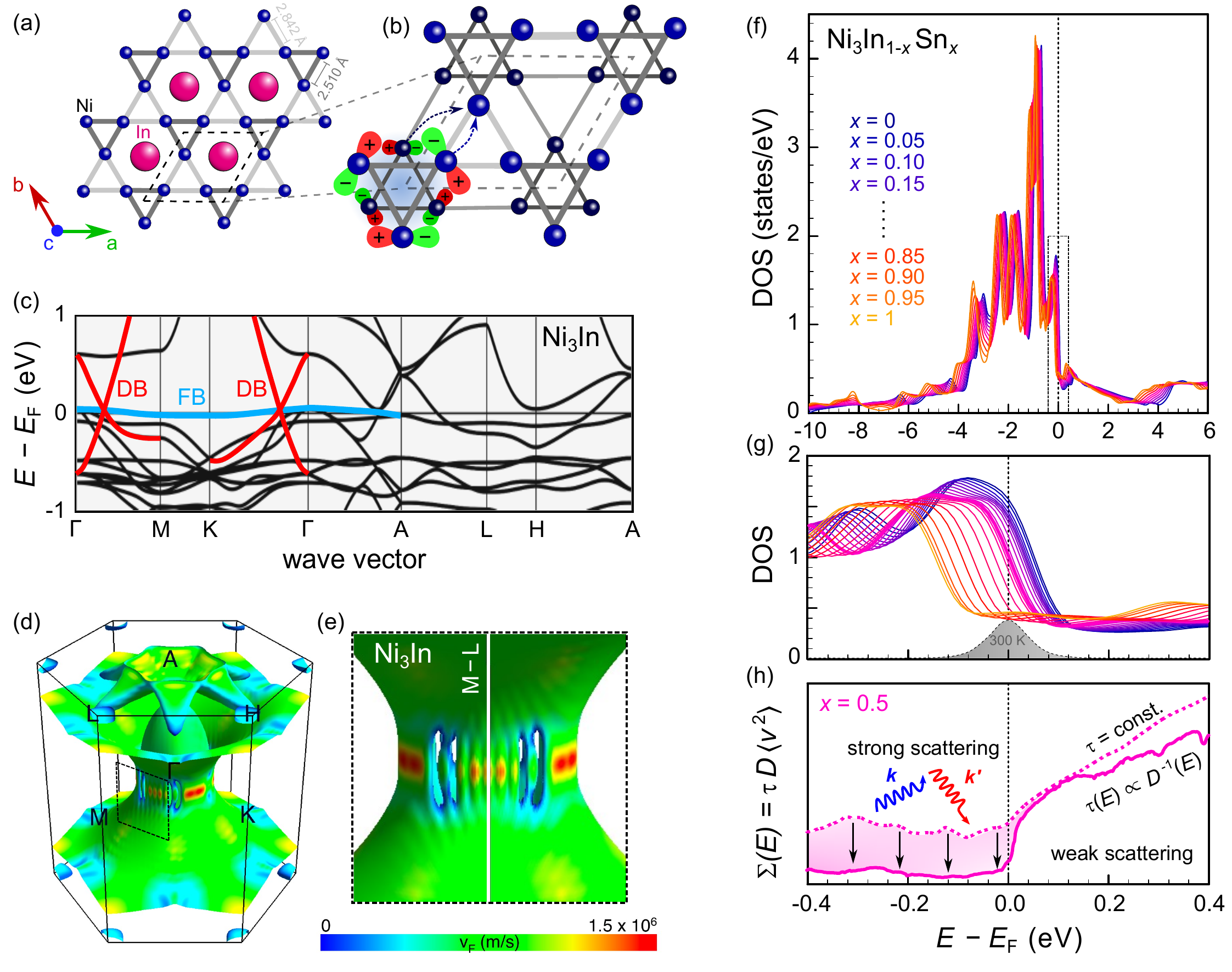}
	\caption{Electronic structure of \ce{Ni3In}-type kagome metals. (a),(b) Top view of kagome lattice planes with triangular motifs formed by Ni atoms in \ce{Ni3(In,Sn)}. The red and green lobes in (b) indicate phases with opposite signs for the $d_{xz}$ wave functions. The alternating phase pattern of the $d_{xz}$-orbital texture between different layers is predicted \cite{ye2024hopping} to lead to destructive interference among inter- and intra-layer hopping paths (arrows) enforcing spatial localization of electronic states at the triangular plaquettes (blue shaded area). (c) Electronic band structure of \ce{Ni3In} hosting flat bands (blue color) and highly dispersive Dirac-like bands (red color) at the Fermi energy $E_\text{F}$. (d) Fermi surface of \ce{Ni3In} with false colors indicating respective contributions of the Fermi velocity. (e) Anomalies in the Fermi surface in the plane spanned by $\Gamma-$\,M and $\Gamma-$\,K directions, \textit{cf.} dashed box in (d). (f),(g) Densities of states (DOS) of \ce{Ni3In_{1-x}Sn_x} for various alloy concentrations. The position of $E_\text{F}$ can be fine-tuned via In/Sn substitution. (h) Electronic transport function $\Sigma(E)\propto \tau(E)\propto D^{-1}(E)$. As the scattering phase space increases with larger DOS, the energy-dependent conductivity $\Sigma(E)$ is suppressed due to strong interband scattering below $E_\text{F}$ and remains large (weak interband scattering) above $E_\text{F}$ in \ce{Ni3In_{0.5}Sn_{0.5}}. 
}
	\label{Fig1}
\end{figure*}

Unfortunately, most ordinary metals show vanishingly small Seebeck coefficients of a few microvolts per kelvin or less due to the lack of a strong energy dependence in the electronic transport distribution function $\Sigma(E)=D(E)\,\langle v(E)^2\rangle \tau(E)$; here, $D(E)$ is the electronic density of states (DOS), $\langle v(E)^2\rangle$ the mean square of the group velocity and $\tau(E)$ the electron relaxation time. In the first approximation, the Mott formula yields $S\propto -\left(\partial\,\text{ln}\,\Sigma(E)/\partial E\right)\vert_{E=E_\text{F}}$. Nonetheless, in certain metallic systems, such as heavy fermion compounds \cite{gambino1973anomalously,rowe2002electrical} or some transition metal alloys \cite{garmroudi2023high}, $\tau(E)$ can acquire a strong energy dependence due to electronic interband scattering, yielding sizeable values of $S$, $PF$ and even $zT$. For example, we recently discovered record-high power factors in simple binary \ce{Ni_xAu_{1-x}} alloys, many times larger than in any other material reported so far above room temperature \cite{garmroudi2023high}. Nonetheless, the high cost of Au as well as the metastable nature of the Ni-Au system pose challenges for practical applications, motivating efforts to screen other metallic systems with similar electronic structures but more abundant and cheaper elements.

Here, we perform a thorough theoretical and experimental study of \ce{Ni3In}-based kagome metals, which have recently been proposed as topological flat-band materials hosting a combination of extremely flat and highly dispersive Dirac-like dispersions at the Fermi energy \cite{ye2024hopping}. The pecularities of the kagome lattice planes (see Fig.\,\ref{Fig1}(a),(b)) are known to give rise to effects such as magnetic frustration in the context of antiferromagnetic interactions \cite{balents2010spin} as well as various exotic quantum phases \cite{han2012fractionalized,
xu2015intrinsic,ye2018massive}. Crucially, they also play a key role in the band flattening driven by destructive phase interference among electronic hopping paths \cite{kang2020topological,kang2020dirac,ye2024hopping}. It has been suggested that such lattice-geometry-enforced band flattening in \ce{Ni3In} may give rise to non-Fermi-liquid behavior, quantum criticality and further provide access to many-body phenomena which integrate correlation with topology \cite{ye2024hopping}. Within this study, we unveil the potential of engineering such type of electronic structures for thermoelectrics and experimentally realize high power factors up to $PF\approx 5$\,mW\,m$^{-1}$\,K$^{-2}$ at room temperature. Apart from the obvious choice of designing functional electronic materials based on the band dispersion and bandwidth inherent to the atomic orbitals ($s$ and $p$ orbitals being less localized than $d$ and $f$ orbitals), our work underscores frustrated lattice geometry as a pivotal tuning knob for designing scattering-tuned metallic thermoelectrics.

\section{Results and Discussion}
In Fig.\,\ref{Fig1}, we summarize the electronic structure of \ce{Ni3In}. Fig.\,\ref{Fig1}(a),(b) shows the kagome lattice formed in the $ab$ planes of this system by Ni atoms. The mismatch in the bond length of $d=2.510\,$\AA\, \textit{vs.} $d^*=2.842\,$\AA\, (light \textit{vs.} dark grey) results in a so-called breathing kagome lattice with AB stacked bilayers (see Fig.\,\ref{Fig1}(b)). The alternating sign pattern of the $d_{xz}$ orbital texture across bilayers results in destructive phase interference among inter- and intra-layer hopping pathways, yielding spatially localized electronic states around the triangular Ni plaquettes (blue shaded area Fig.\,\ref{Fig1}(b)) \cite{ye2024hopping}. The band structure of \ce{Ni3In}, calculated by density functional theory (DFT) methods is shown in Fig.\,\ref{Fig1}(c). The most striking features are two dispersive Dirac-like bands (DB) crossing the Fermi energy $E_\text{F}$ along the $\Gamma-$\,M and $\Gamma-$\,K line together with a remarkably flat, sinusoidal band. The latter disperses only very slightly (bandwidth $W\lesssim 80\,$meV) in a wave-like manner and crosses $E_\text{F}$ two times. This double crossing creates peculiar features in the Fermi surface (FS) shown in Fig.\,\ref{Fig1}(d),(e), namely vertical cuts, which correspond to electron- and hole-like features at different parts in the Brillouin zone. In other words, charge carriers associated with the flat-band states should behave as electron-like at the $\Gamma$ point, whereas heavy hole-like quasiparticle excitations are expected closer towards the K and M point. The separate sheets of the FS along with phonon dispersions and electron-phonon spectral functions as well as computational details are presented in the Supplemental Materials~\cite{supplemental}. Here, we focus on the variation  of the Fermi velocity $v_\text{F}$ along the FS, which is plotted in color in Fig.\,\ref{Fig1}(d). A magnification of the $v_\text{F}$ anomalies is depicted in Fig.\,\ref{Fig1}(e). It can be seen that much larger values of $v_\text{F}$ are found along $\Gamma-$\,M and $\Gamma-$\,K, corresponding to the highly dispersive Dirac-like bands in Fig.\,\ref{Fig1}(c). Additionally, right next to these maxima, $v_\text{F}$ drops rapidly and vertical cuts of low group velocity are formed that are attributed to the aforementioned flat and sinusoidal band.

\begin{figure}[t!]
\newcommand{\setwidth}{0.45}
			\centering
			\hspace*{-0.25cm}
		\includegraphics[width=0.45\textwidth]{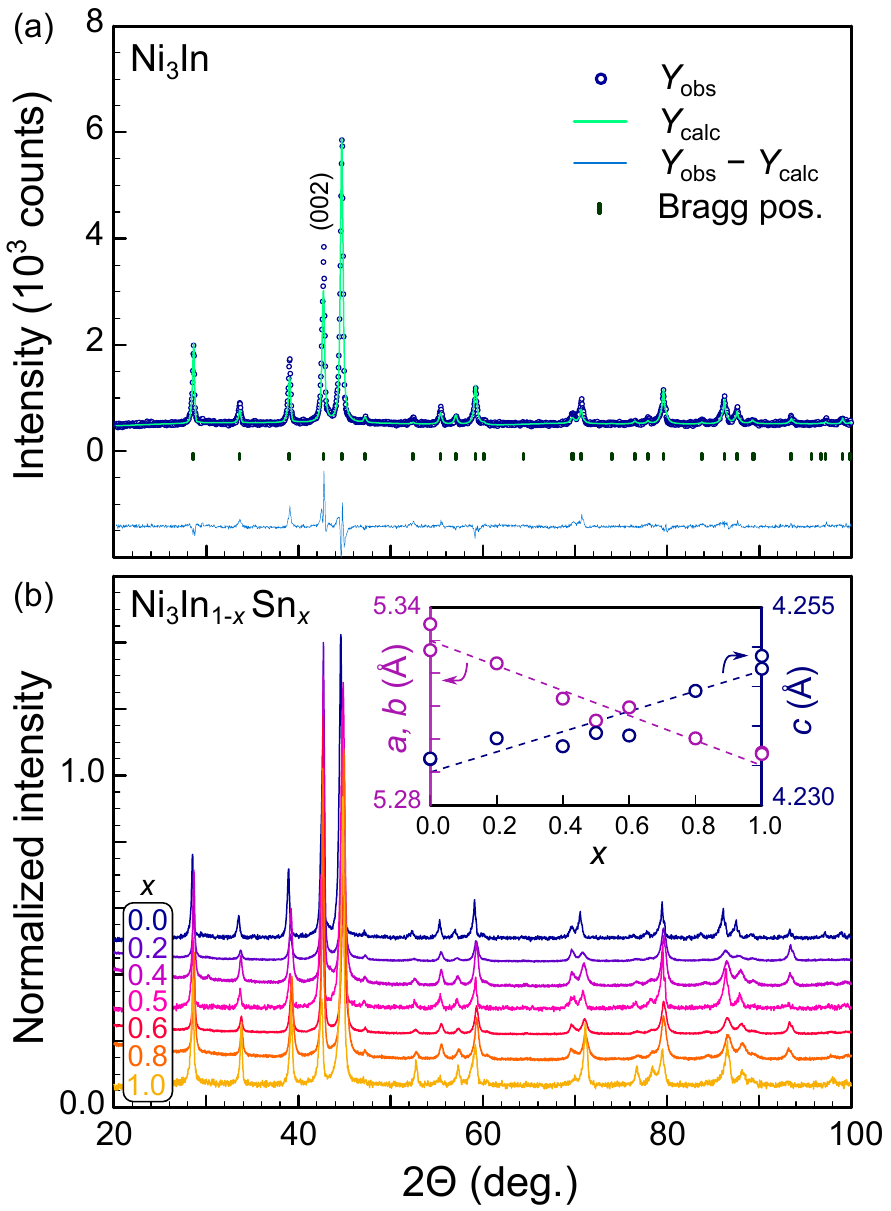}
	\caption{Structural properties of \ce{Ni3In_{1-x}Sn_x}. (a) X-ray powder diffraction pattern of phase-pure \ce{Ni3In} with Rietveld refinement (solid line) and Bragg positions. (b) Experimental patterns for all investigated \ce{Ni3In_{1-x}Sn_x} samples. Inset shows lattice parameters of as a function of Sn concentration.
} 
	\label{Fig2}
\end{figure}

To control the position of the Ni-$3d$ flat-band edge with respect to $E_\text{F}$, we also investigated the effect of Sn substitution at the In site in \ce{Ni3In_{1-x}Sn_x} since \ce{Ni3Sn} exists as an isostructural variant and a full solid solution is possible. Fig.\,\ref{Fig1}(f),(g) shows the alloy-averaged densities of states of \ce{Ni3In_{1-x}Sn_x} calculated within the framework of the Kohn-Korringa-Rostoker method and the coherent potential approximation (KKR-CPA). This method accounts for alloy disorder due to the substitution of Sn atoms at the In site and has been previously employed successfully for calculating the electronic structure of many thermoelectric materials such as Heusler alloys \cite{parzer2022high,
garmroudi2022anderson}, binary transition metal alloys \cite{butler1984calculated,
banhart1995first,wiendlocha2018thermopower,
garmroudi2023high} but also semiconductor systems such as PbTe \cite{heremans2012resonant}. Fig.\,\ref{Fig1}(f) indicates that the DOS of \ce{Ni3In_{1-x}Sn_x} features a steep edge at $E_\text{F}$, mainly arising from the flat band, \textit{i.\,e.}, Ni-$d_{xz}$ orbitals localized at the triangular plaquettes \cite{ye2024hopping}. Crucially, it can be seen that the position of this $d$ edge can be tuned by $x$ with respect to $E_\text{F}$ in the energy range relevant to thermoelectric transport (Fig.\,\ref{Fig1}(g)). The grey shaded area represents the derivative of the Fermi-Dirac distribution function at 300\,K, which marks the relevant energy range for electronic transport at room temperature. With increasing Sn concentration in \ce{Ni3In_{1-x}Sn_x}, $E_\text{F}$ is shifted in a rigid-band-like manner by almost 0.2\,eV from $x=0$ to $x=1$. 

Charge carriers from the underlying dispersive Dirac-like bands can scatter into the flat-band states via interband scattering. Due to the rapid variation of $D(E)$, the scattering phase space exhibits a strong energy dependence. The large phase space for dispersive conduction carriers scattering into localized flat-band Ni-$d$ states yields a reduction of $\Sigma(E)$ at $E<E_\text{F}$, while the contribution of electronic excitations to the conductivity remains high at $E>E_\text{F}$, thereby promoting a pronounced electron--hole asymmetry. This imbalance is shown in Fig.\,\ref{Fig1}(h) for \ce{Ni3In_{0.5}Sn_{0.5}}, for which we experimentally observe the highest TE performance.

\begin{figure*}[t]
\newcommand{\setwidth}{0.45}
			\centering
			\hspace*{-0.25cm}
		\includegraphics[width=0.95\textwidth]{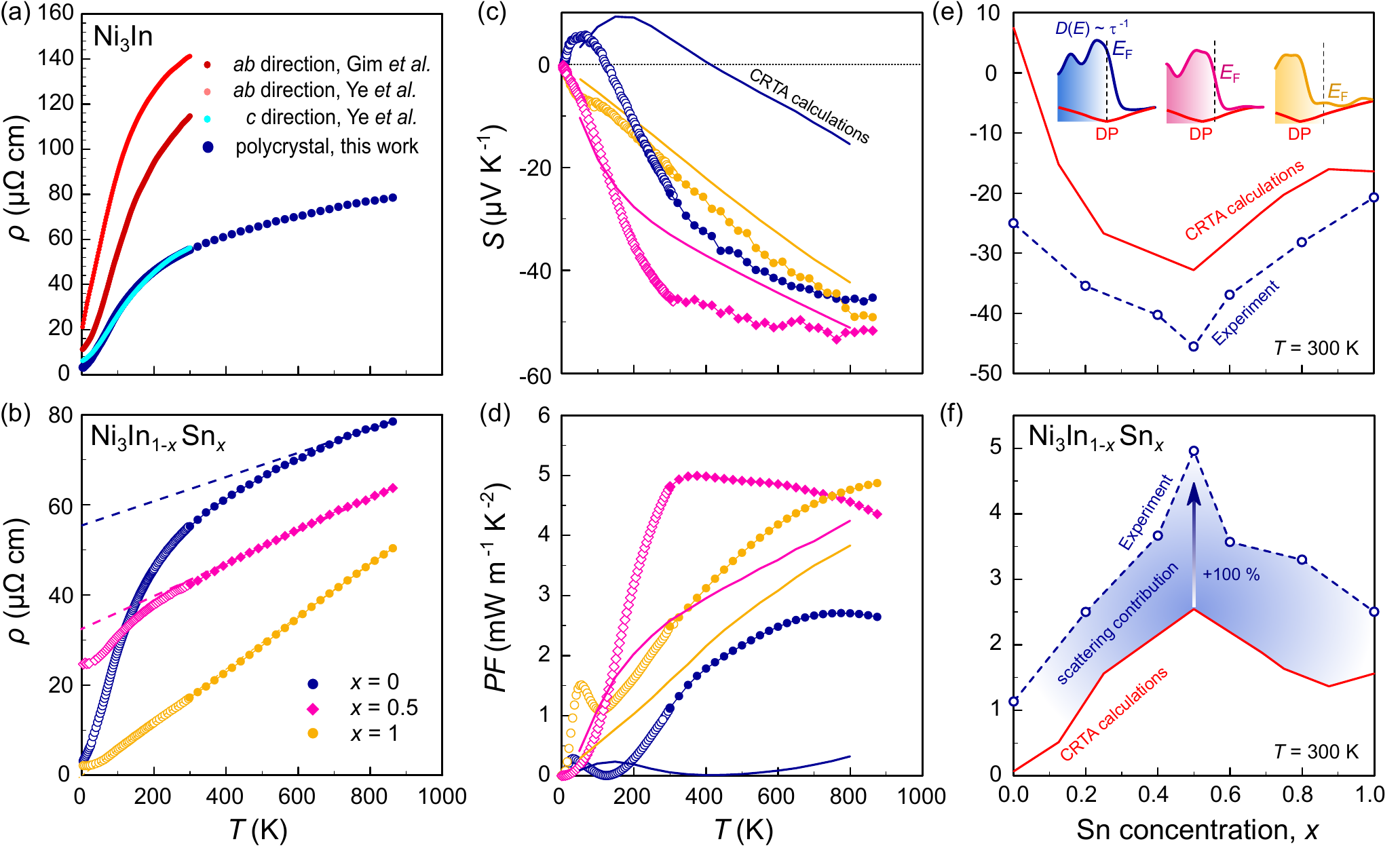}
	\caption{Thermoelectric properties of \ce{Ni3In_{1-x}Sn_x}. (a) Temperature-dependent electrical resistivity of the pristine \ce{Ni3In} from this work compared to single crystal data from Refs.\,\cite{ye2024hopping} and \cite{gim2023fingerprints} measured parallel and perpendicular to the kagome lattice planes. (b) Temperature-dependent electrical resistivity, (c) Seebeck coefficient and (d) power factor of \ce{Ni3In_{1-x}Sn_x} ($x=0,\,0.5,\,1$) in a very broad temperature range 4\,--\,860\,K. Solid lines are theoretical calculations in the constant relaxation time approximation (CRTA) using DFT electronic structures. The Seebeck coefficient is calculated without any free parameters. The power factor is calculated by making use of the experimental resistivity. (e) Experimental room temperature Seebeck coefficient and (f) power factor as a function of the Sn concentration in \ce{Ni3In_{1-x}Sn_x} compared to CRTA results. The inset in (e) shows the variation of the flat-band edge (corresponding to the scattering rate) around the Fermi energy for \ce{Ni3In} (blue), \ce{Ni3In_{0.5}Sn_{0.5}} (pink) and \ce{Ni3Sn} (yellow). The DOS of the underlying Dirac point responsible for the composition-dependent CRTA results is sketched in red color.
	}
	\label{Fig3}
\end{figure*}

Since large samples are an essential requirement for conducting high-temperature thermoelectric measurements, we prepared polycrystalline samples in the course of this study. In Fig.\,\ref{Fig2}(a), a powder x-ray diffraction pattern and the corresponding Rietveld refinement of phase-pure \ce{Ni3In} (hexagonal phase, space group $P6_3/mmc$) are presented. The diffraction patterns of all samples synthesized and investigated in this work are shown in Fig.\,\ref{Fig2}(b) together with the evolution of the lattice parameters obtained from the Rietveld refinements (see inset). The $a$,\,$b$ lattice parameter in the kagome plane decreases while $c$ increases in the cross-plane direction with increasing Sn concentration. The linear behavior further confirms the full solubility of Sn in \ce{Ni3In_{1-x}Sn_x}. 


\begin{figure}[tbh]
\newcommand{\setwidth}{0.45}
			\centering
			\hspace*{-0.25cm}
		\includegraphics[width=0.45\textwidth]{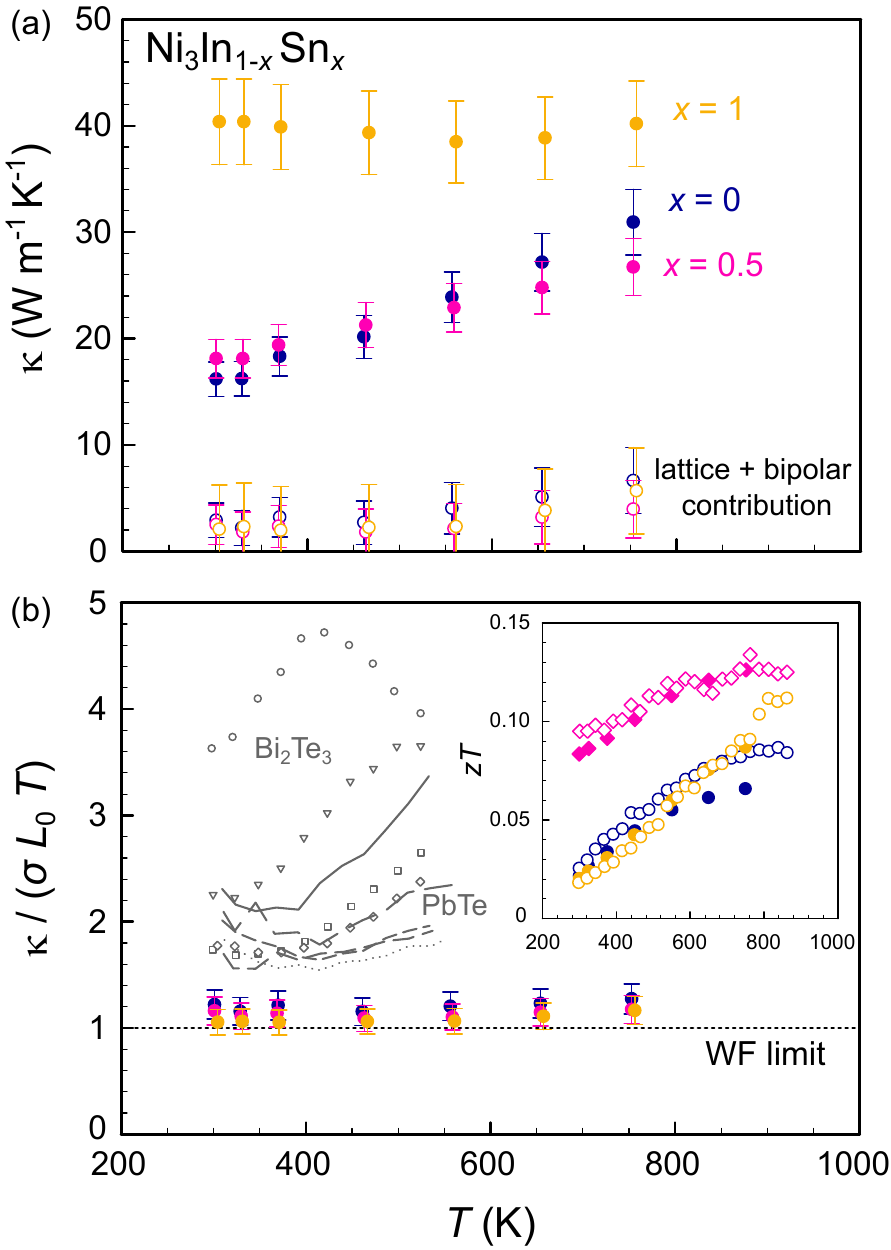}
	\caption{Thermal transport at the Wiedemann-Franz limit in \ce{Ni3In_{1-x}Sn_x}. (a) Total thermal conductivity of \ce{Ni3In_{1-x}Sn_x} with $x=0,\,0.5,\,1$ together with lattice and bipolar contributions (open symbols). (b) Relative deviation of thermal transport from the Wiedemann-Franz limit $L/L_0=\kappa/(\sigma L_0 T)$ illustrates irrelevance of phononic heat transport in metallic \ce{Ni3In_{1-x}Sn_x}. Archetypal chalcogenide semiconductors with various doping concentrations (grey symbols \ce{Bi2Te3}:I \cite{witting2019thermoelectric} and lines \ce{PbTe}:Na \cite{wang2018sodium}) are shown as comparison for thermal transport in state-of-the-art thermoelectrics. Inset shows $zT$ of \ce{Ni3In_{1-x}Sn_x} obtained from experimental data of $S$, $\rho$ and $\kappa$ as well as estimations solely considering the experimental Seebeck coefficient $zT=S^2/L$ (open symbols).} 
	\label{Fig4}
\end{figure}

Fig.\,\ref{Fig3} summarizes the TE transport properties of polycrystalline \ce{Ni3In_{1-x}Sn_x}. The highly anisotropic electronic structure is reflected in a significant anisotropy in electrical resistivity, with measurements within the kagome planes of \ce{Ni3In} showing much larger values compared to those in the cross-plane direction \cite{ye2024hopping,gim2023fingerprints}. 
The results reported for the $c$ direction by Ye \textit{et al.} \cite{ye2024hopping} are in very good agreement with our work, suggesting that either our polycrystal is textured or mainly the $c$ direction is measured (see Fig.\,\ref{Fig3}(a)). The former was ruled out by measuring $\rho$ in different directions (see Fig.\,S2), confirming the isotropy of our samples. 
Note that the reported residual resistivity ratio (RRR) for single crystals measured in the $ab$ plane by Ye \textit{et al.} \cite{ye2024hopping} and Gim \textit{et al.} \cite{gim2023fingerprints} is 6.2 and 10.2, respectively, which is even smaller than RRR$=17.8$ reported here for \ce{Ni3In}, possibly related to antisite defects in the single crystals and the fact that \ce{Ni3In} is not a line compound. This showcases the high quality as well as the negligible effect of grain boundary scattering in our polycrystalline samples.

Fig.\,\ref{Fig3}(b) compares the temperature-dependent electrical resistivity for Sn-substituted \ce{Ni3In_{1-x}Sn_x} with $x=0,\,0.5,\,1$. The temperature dependence of additional samples ($x=0.2,\,0.4,\,0.6,\,0.8$) is shown in the SM. Despite the substitutional disorder by alloying Sn on the In site, $\rho(T)$ decreases with increasing $x$ at high temperatures ($T\gtrsim 100\,$K). This is very much consistent with DFT calculations (Fig.\,\ref{Fig1}(g)) showing that $E_\text{F}$ is pushed above the flat band leading to a decrease of the overall scattering rate. For \ce{Ni3Sn}, $\rho$ reaches exceptionally small values $<20\,\mu \Omega$\,cm at 300\,K, comparable even to some elemental transition metals, manifesting the clean, metallic nature of this system. A noticeable feature of In-rich samples is the gradual change in slope of $\rho(T)$ (see the dashed lines in Fig.\,\ref{Fig3}(b)). For simple metals, a linear-in-temperature scattering rate is expected due to dominant acoustic phonon scattering at high temperatures (see \textit{e.\,g.} \ce{Ni3Sn}). The change in slope, most noticeable for \ce{Ni3In}, stems from the edge of the flat band becoming thermally activated at high temperatures, which leads to the chemical potential being shifted up to higher energies as temperature increases. The deviation from high-temperature linear behavior is not only shifted to lower temperatures in \ce{Ni3In_{0.5}Sn_{0.5}} but is also much more abrupt, in line with $E_\text{F}$ being closer to the flat-band edge (see Fig.\,\ref{Fig1}(g),(h)). Similar conclusions can be drawn from thermoelectric transport measurements discussed below.

Figs.\,\ref{Fig3}c,d display the temperature-dependent Seebeck coefficient and the power factor $PF=S^2/\rho$. We compare our experimental data to theoretical calculations (solid lines) based on the commonly utilized constant relaxation time approximation (CRTA), derived within the framework of the Boltzmann transport formalism using electronic structures obtained from our DFT calculations performed using the {\sc WIEN2k} package~\cite{wien2k,wien2k2020} and the {\sc BoltzTrap} code~\cite{boltztrap}. The anisotropic thermopower was averaged according to the parallel grain model of the polycrystalline material~\cite{snse}
\begin{equation}
S=\dfrac{S_{xx}\sigma_{xx}+S_{yy}\sigma_{yy}+S_{zz}\sigma_{zz}}{\sigma_{xx}+\sigma_{yy}+\sigma_{zz}}.
    \label{eq:seebeck-eff}
\end{equation}
Here, conductivities $\sigma$, along the diagonal directions, are used as weights. More details of these calculations are given in the Supplemental Material~\cite{supplemental}.

The scattering rate drops out in the Boltzmann expression for the Seebeck coefficient, leaving no free parameters in the model; the resistivity is proportional to the scattering rate. The power factor was calculated by using the theoretically calculated Seebeck coefficient and experimental resistivity values.
For \ce{Ni3Sn}, where the upper edge of the flat band is far below $E_\text{F}$ (weak scattering), $S(T)$ can be well described in the CRTA framework. However, for samples closer to the flat-band edge, interband scattering of mobile carriers scattering into flat-band states becomes important. This is not taken into account in the CRTA and leads to an additional negative contribution to the Seebeck coefficient. Here, the larger DOS below $E_\text{F}$ heavily impedes hole transport while electrons at $E>E_\text{F}$ remain highly mobile. Since interband transitions are also mediated by phonons, especially in pristine \ce{Ni3In}, the additional Seebeck contribution becomes progressively smaller towards lower temperatures, where the phonon wave vector is too small to allow for extensive interband scattering. 
When $E_\text{F}$ is close to the flat-band edge, the chemical potential gets shifted towards higher energies as a function of temperature, causing a saturation of $S(T)$. Once again, the saturation starts at lower temperatures in \ce{Ni3In_{0.5}Sn_{0.5}} compared to \ce{Ni3In}, in line with the aforementioned trend of $\rho(T)$.

Figs.\,\ref{Fig3}d emphasizes that the power factor is greatly enhanced due to this interband scattering mechanism. While CRTA calculations, ignoring interband transitions, predict a vanishingly small power factor for \ce{Ni3In}, $PF$ reaches up to 2.7\,mW\,m$^{-1}$\,K$^{-2}$ in our experiments and even goes up to 5\,mW\,m$^{-1}$\,K$^{-2}$ at room temperature for the Sn-substituted sample, twice as large as predicted by calculations. Composition-dependent trends for both $S$ and $PF$ at room temperature are presented in Figs.\,\ref{Fig3}(e),(f). When interband scattering is neglected, \ce{Ni3In} shows a negligible Seebeck effect due to the fact that $E_\text{F}$ is situated close to the Dirac points with perfect electron-hole symmetry (see Fig.\,\ref{Fig1}(c) and inset of Fig.\,\ref{Fig3}(e)). The rather localized flat-band states do not actively contribute to thermoelectric transport due to their high effective mass and low conductivity weight. $S$ and $PF$ peak around $x=0.5$ where the effective carrier concentration of Dirac-like carriers is optimized, in agreement with our experimental results. However, the absolute values are sizeably larger for the latter owing to the increased electron--hole asymmetry associated with the energy-dependent scattering, which is neglected in calculations assuming a constant scattering time.

To investigate heat transport and experimentally evaluate $zT$ of \ce{Ni3In_{1-x}Sn_x}, we measured the temperature-dependent thermal conductivity $\kappa(T)$ for three selected samples with $x=0,\,0.5,\,1$ (see Fig.\,\ref{Fig4}(a)). It can be seen that lattice (and bipolar) contributions make up an almost insignificant portion of overall thermal conductivity, which means that these compounds are good metals also in terms of the Wiedemann-Franz law $\kappa\approx \kappa_e=L_0\sigma T$, with $L_0=\frac{\pi^2}{3}\left(\frac{k_\text{B}}{e}\right)^2$. This is further illustrated in Fig.\,\ref{Fig4}(b), where we plot the relative deviation of our samples from the Sommerfeld value of the Lorenz number $L_0$, which is reached for $\kappa_l\rightarrow 0$ or $\kappa_e\rightarrow \infty$. All of our investigated samples are very close to this Wiedemann-Franz limit, meaning that they realize ideal heat transport for TE materials (electron-dominated instead of lattice-dominated). On the other hand, even the best semiconducting systems, such as chalcogenides based on \ce{Bi3Te3} or \ce{PbTe} with notoriously small $\kappa_l$ are significantly away from this limit. The inset in Fig.\,\ref{Fig4}(b) shows $zT$ calculated from experimental data of $S$, $\rho$ and $\kappa$ (filled symbols) as well as estimated solely from the Seebeck coefficient via $zT=S^2/L$ (open symbols). Indeed, the latter agrees very well within the error bar of our experimental setup ($\approx 20\,\%$ for $zT$) showcasing that $zT$ only depends on the Seebeck coefficient in these kagome metals. 

\begin{figure}[t]
\newcommand{\setwidth}{0.45}
			\centering
			\hspace*{-0.25cm}
		\includegraphics[width=0.45\textwidth]{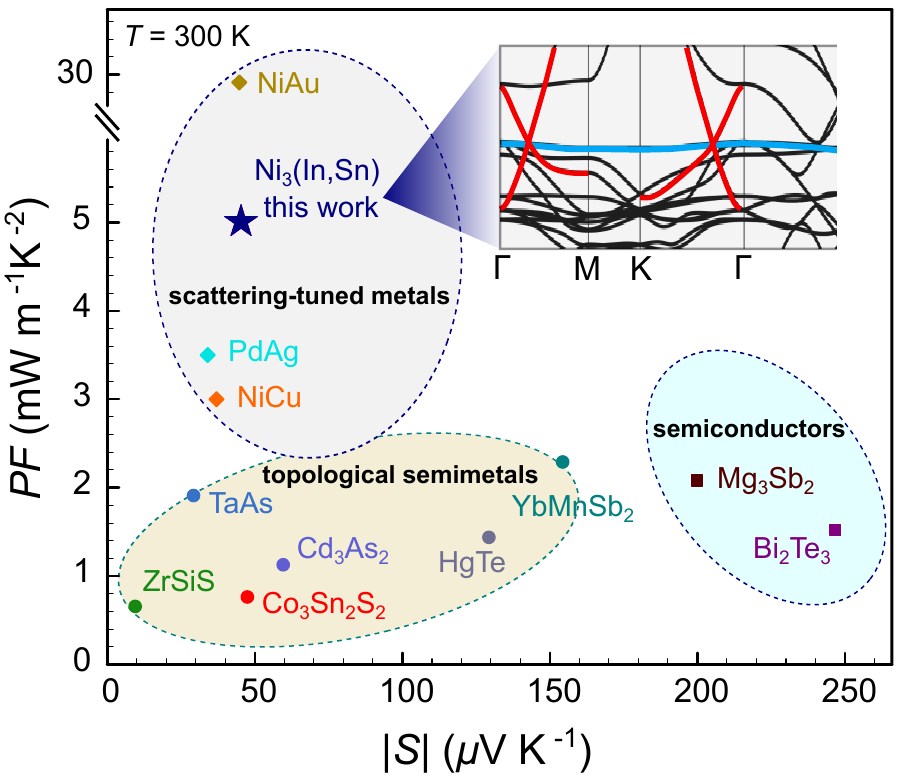}
	\caption{Comparison with other material classes. Power factor as a function of the absolute value of the Seebeck coefficient for topological semimetals \cite{singha2017large,matusiak2017thermoelectric,
xiang2017anisotropic,mangelis2017effect,wang2018magnetic,
markov2018semi,pan2021thermoelectric}, scattering-tuned metals \cite{ho1978thermal,ho1983electrical,
ho1993thermoelectric} and state-of-the-art semiconductors \cite{jeon1991electrical,imasato2020metallic} at $T=300\,$K. Insets show the distinguished electronic structure features in \ce{Ni3In}-type kagome metals responsible for the enhanced Seebeck coefficient and power factor.
}
	\label{Fig5}
\end{figure}

\section{Outlook}
Theoretical studies have pointed out that anomalous sign dependencies of $S$ could manifest in topological insulators, wherein charge carriers scatter from Dirac-like surface states into the bulk bands \cite{takahashi2012thermoelectric,xu2014enhanced,
xu2017topological}. The idea is that when $E_\text{F}$ is placed close to the bulk band edge, charge carriers within the gap are topologically protected from backscattering while being strongly scattered near the band edge owing to boundary-bulk interactions, creating significant variations in $\tau(E)$ around $E_\text{F}$. However, so far, these predictions have neither been validated experimentally nor has an enhancement of $PF$ been successfully realized. In the case of interband scattering between Dirac-like and topological flat-band states in \ce{Ni3In_{1-x}Sn_x} kagome metals, a similar scenario emerges. Notably, this phenomenon occurs as an intrinsic bulk effect, offering a distinct advantage over the aforementioned scenario \cite{takahashi2012thermoelectric,xu2014enhanced,
xu2017topological} that would require sophisticated nanostructuring, and that on a large scale for achieving the necessary throughput for actual applications.

Topological semimetals, such as Dirac and Weyl semimetals, typically exhibit rather low carrier concentrations, comparable or only slightly larger than those of narrow-gap semiconductors. Therefore, most semimetals also do not come close to the Wiedemann-Franz limit (electron-dominated heat transport) and require substantial tuning with respect to their lattice thermal conductivity, offering no distinct advantages over currently investigated semiconductors. Fig.\,\ref{Fig5} compares the room temperature power factor and the absolute Seebeck coefficient of various topological semimetals like ZrSiS \cite{singha2017large,matusiak2017thermoelectric}, TaAs \cite{xiang2017anisotropic}, \ce{Co3Sn2S2} \cite{mangelis2017effect}, \ce{Cd3As2} \cite{wang2018magnetic}, HgTe \cite{markov2018semi} or recently reported \ce{YbMnSb2} \cite{pan2021thermoelectric} as well as single crystal data of state-of-the-art semiconductors \ce{Bi2Te3} \cite{jeon1991electrical} and \ce{Mg3Sb2} \cite{imasato2020metallic} with those of scattering-tuned metals \cite{ho1978thermal,ho1983electrical,
ho1993thermoelectric}, including the topological flat-band candidate system \ce{Ni3(In,Sn)}. The power factor at room temperature, $PF\approx 5\,$mW\,m$^{-1}$\,K$^{-2}$, reached in \ce{Ni3(In,Sn)} is comparable or even larger than the values that have currently been achieved by most other TE materials. While there have been reported larger room-temperature values of $PF$ in certain systems such as \ce{Fe2VAl}-based full-Heusler \cite{garmroudi2022large,
alleno2023optimization} and \ce{NbFeSb}-based half-Heusler compounds \cite{he2016achieving}, heavy fermion compounds like \ce{YbAl3} \cite{rowe2002electrical} and \ce{CePd3} \cite{gambino1973anomalously} or the recently discovered \ce{Ni_xAu_{1-x}} alloys \cite{garmroudi2023high}, our work underscores the prospect of utilizing interband scattering and provides a proof-of-concept that engineering the interplay of flat and dispersive bands of kagome metals through frustrated lattice geometry can leverage metallic thermoelectrics. This opens an alternative pathway for controlling the bandwidth beyond mere chemical pressure tuning, as has been performed in binary \ce{Ni_xAu_{1-x}} alloys \cite{garmroudi2023high}. Additionally, given that the \ce{Ni3In} structure type exhibits the capability to accommodate a diverse range of elements and compositions, one can anticipate a considerable level of adjustability in the flat-band states and interband scattering potentials through the incorporation of various alloying elements. Thus, there is a high potential for further boosting the thermoelectric properties. 

\section{Conclusion}
In summary, we have shown that high TE power factors up to $5$\,mW\,m$^{-1}$\,K$^{-2}$ at room temperature can be achieved in \ce{Ni3In_{1-x}Sn_x} kagome metals due to strong interband scattering, wherein hole-like charge carriers are scattered from highly dispersive into flat-band states. It has been demonstrated theoretically and experimentally that the position of the flat-band edge with respect to $E_\text{F}$ and thus the energy-dependent scattering can be precisely tuned via Sn substitution on the In site. Our work highlights the potential of engineering topological flat-band states in conjunction with linear Dirac- and Weyl-like dispersions to realize sizeable Seebeck coefficients even in perfect metallic conductors. Apart from conventional TE power generation and refrigeration applications where $zT$ determines the performance and efficiency, high-$PF$ metallic thermoelectrics are desirable for the emergent field of active cooling \cite{adams2019active,
komatsu2021macroscopic,luo2022high}, where hot spots are cooled to near ambient conditions, \textit{e.\,g.} in integrated circuit chips.

\section*{Ackowledgments}
Research in this paper was financially supported by the Japan Science and Technology Agency (JST) program MIRAI, JPMJMI19A1. Xinlin Yan from the Institute of Solid State Physics, TU Wien is acknowledged for fruitful discussions regarding the materials synthesis as well as for assistance with performing Rietveld refinements on x-ray diffraction patterns.


%

\end{document}